\documentclass[aps,preprint,preprintnumbers,amsmath,showpacs,amssymb,nofootinbib]{revtex4}
\usepackage{amssymb}
\usepackage{amsmath}
\usepackage{bm}
\usepackage{graphicx}
\draft
\begin{document}
\title{Dirac equation in a de Sitter expansion for massive neutrinos from modern Kaluza-Klein theory.}
\author{$^{1}$ Pablo Alejandro S\'anchez\footnote{pabsan@mdp.edu.ar}, Mariano Anabitarte\footnote{anabitar@mdp.edu.ar} and Mauricio
Bellini\footnote{mbellini@mdp.edu.ar}}
\address{$^{1}$ Departamento de F\'{\i}sica, Facultad de Ciencias Exactas y
Naturales, \\
Universidad Nacional de Mar del Plata, Funes 3350, (7600) Mar del
Plata, Argentina. \\ \\
$^{2}$ Instituto de Investigaciones F\'{\i}sicas de Mar del Plata
(IFIMAR), Consejo Nacional de Investigaciones Cient\'{\i}ficas y
T\'ecnicas (CONICET), Argentina.}
\begin{abstract}
Using the modern Kaluza-Klein theory of gravity (or the Induced
Matter theory), we study the Dirac equation for massive neutrinos
on a de Sitter background metric from a 5D Riemann-flat (and hence
Ricci-flat) extended de Sitter metric, on which is defined the
vacuum for test massless $1/2$-spin neutral fields minimally
coupled to gravity and free of any other interactions. We obtain
that the effective 4D masses of the neutrinos can only take three
possible values, which are related to the (static) foliation of
the fifth and noncompact extra dimension.
\end{abstract}
\maketitle

\section{Introduction}

In models based on supergravity \cite{1}, it has been pointed out
the existence of some light particles whose interactions are
suppressed at scales close to $M = M_P/(8\pi)\simeq 2.4\times
10^{18}$ GeV. Such particles have nothing to do with the collider
experiments, but may affect the standard scenario of big-bang
cosmology \cite{2,3,4}. The gravitino, which is the gauge field
associated with local supersymmetry (SUSY), is one of the weakly
interacting particles in supergravity models, and we expect the
mass of the gravitino to be of the order of the typical
SUSY-breaking scale. When the gravitino decays into a neutrino and
a sneutrino, the emitted high energy neutrinos scatter off the
background neutrinos and produce charged leptons (mainly electrons
and positrons), which cause an electro-magnetic cascades and
produce many soft photons. Hence, the propagation of neutrinos
should be very important during inflation. Inflationary cosmology
can be recovered from a 5D vacuum\cite{nos,nosb,nosc}. The
inflationary theory is very consistent with current observations
of the temperature anisotropy of the Cosmic Microwave Background
(CMB)\cite{5}. The most popular model of supercooled inflation is
chaotic inflation\cite{6}. In this model the expansion of the
universe is driven by a single scalar field called inflaton. At
some initial epoch, presumably the Planck scale, the scalar field
is roughly homogeneous and dominates the energy density, which
remains almost constant during all the inflationary epoch.

On the other hand inflation can be recovered from the
Campbell-Magaard
theorem\cite{campbell,c1,campbellb,campbellc,campbelld}, which
serves as a ladder to go between manifolds whose dimensionality
differs by one. This theorem, which is valid in any number of
dimensions, implies that every solution of the 4D Einstein
equations with arbitrary energy momentum tensor can be embedded,
at least locally, in a solution of the 5D Einstein field equations
in vacuum. Because of this, the stress-energy may be a 4D
manifestation of the embedding geometry. Physically, the
background metric there employed describes a 5D extension of an
usual de Sitter spacetime, which is the 4D spacetime that
describes an inflationary expansion.

In this Letter we study the Dirac equation for 4D massive
neutrinos in a de Sitter expansion using Modern Kaluza-Klein
theory of gravity (or Induced Matter theory)\cite{wo,we}. In this
theory the 5D massless $1/2$-spin test fields are considered free
from interactions and minimally coupled to gravity on a 5D
Ricci-flat metric in which we define the physical vacuum. Our
approach is something different (but complementary) than the
studied by Wesson in \cite{ww}, because we make a detailed study
of the geometrical spinor properties of 5D vector fields that,
once we make a static foliation on the fifth coordinate, can be
considered as 4D massive neutrinos.

\section{The 5D Clifford algebra and spinors in 5D}

We consider the Ponce de Le\'on metric\cite{PdL}
\begin{equation}\label{a1}
dS^2=\left(\frac{\psi}{\psi_0}\right)^2\left[dt^2-e^{2t/\psi_0}dR^2\right]-d\psi^2.
\end{equation}
The resulting 4D hypersurface after making $\psi=\psi_0$ describes
a de Sitter spacetime. From the relativistic point of view an
observer moving with the penta-velocity $U_\psi=0$, will be moving
on a spacetime that describes a de Sitter expansion which has a
scalar curvature $^{(4)} R=12/\psi^2_0=12\,H^2_0$, such that the
Hubble parameter is defined by the foliation $H_0=\psi_0^{-1}$. If
we foliate $\psi=\psi_0$, we get the effective 4D metric
 \begin{equation}
    dS^2\rightarrow ds^2=dt^2-e^{2H_0t}d\vec{R}^2,\label{24}
 \end{equation}
which describes a 3D spatially flat, isotropic and homogeneous de
Sitter expanding universe with a constant Hubble parameter $H_0$.

To define a 5D vacuum we shall consider a Lagrangian for a
massless 5D spinor field minimally coupled to gravity (we shall
consider $\hbar=c=1$)
\begin{equation}\label{ac}
L = \frac{1}{2} \left[ \bar{\Psi} \gamma^A \left(\nabla_A
\Psi\right) - \left( \nabla_A \bar{\Psi}\right)
\gamma^A\Psi\right] + \frac{R}{2\bar{K}},
\end{equation}
where $\bar{K}={8\pi G}$ and $\gamma^A$ are the Dirac matrices
which satisfy
\begin{equation}
\left\{ \gamma^A, \gamma^B\right\} = 2 g^{AB} \;\mathbb{I},
\end{equation}
such that the covariant derivative of the spinor $\Psi$ on
(\ref{a1}) is defined in the following form:
\begin{equation}\label{na}
\nabla_A \Psi = \left( \partial_A +  \Gamma_A \right) \Psi,
\end{equation}
and the spin connection is given by
\begin{equation}\label{ga}
\Gamma_A =\frac{1}{8} \left[ \gamma^b, \gamma^c\right]
e_{\,\,b}^B\, \nabla_A\left[ e_{cB}\right],
\end{equation}
being $\nabla_A\left[ e_{cB}\right]=\partial_A e_{cB} -
\Gamma^D_{AB} e_{cD}$ the covariant derivative of the five-bein
$e_B^c$ (the symbol $\partial_A$ denotes the partial derivative
with respect to $x^A$ and $\eta_{ab}= g_{AB} e^A_{\,\,a}
e^B_{\,\,b}$ denotes the 5D Minkowski spacetime in Cartesian
coordinates), which relates the extended 5D de Sitter metric
(\ref{a1}) with the 5D Minkowski spacetime written in Cartesian
coordinates: $dS^2= dt^2 - (dx^2+dy^2+dz^2)-d\psi^2$
\begin{equation}
e_B^c= \left( \begin{array}{lllll} \left(\frac{\psi}{\psi_0}\right)   & 0 & 0 & 0 & 0 \\
0 & \left(\frac{\psi}{\psi_0}\right) e^{t/\psi_0} & 0 & 0
&  0 \\
0 & 0 & \left(\frac{\psi}{\psi_0}\right) e^{t/\psi_0}
& 0 \\
0 & 0 & 0 & \left(\frac{\psi}{\psi_0}\right) e^{t/\psi_0} & 0 \\
0 & 0 & 0 & 0 &  1\end{array} \right).
\end{equation}
The Dirac matrices $\gamma^a$ are represented in an Euclidean
space instead of a Lorentzian space, and are described by the
algebra\cite{grov,camporesi}: $\left\{ \gamma^a,\gamma^b\right\} =
2 \eta^{ab} \mathbb{I}$
\begin{eqnarray}
&& \gamma^0= \left( \begin{array}{llll} 1 & 0 & 0 & 0  \\
0 & 1 & 0 &  0 \\
0 & 0 & -1 & 0 \\
0 & 0 & 0 & -1
\end{array} \right) = \left(\begin{array}{ll}  \mathbb{I} & 0 \\
0 &  \mathbb{I}\ \end{array} \right),\qquad
\gamma^1= \left( \begin{array}{llll} 0 & 0 & 0 & 1 \\
0 & 0 & 1 & 0 \\
0 & -1 & 0 & 0 \\
-1 & 0 & 0 & 0 \end{array} \right) = \left(\begin{array}{ll} 0 &  \sigma^1 \\
- \sigma^1 & 0  \end{array} \right),  \nonumber \\
&& \gamma^2= \left( \begin{array}{llll} 0 & 0 & 0 & -i \\
0 & 0 & i & 0 \\
0 & i & 0 & 0 \\
-i & 0 & 0 & 0 \end{array} \right) = \left(\begin{array}{ll} 0 &  \sigma^2 \\
- \sigma^2 & 0  \end{array} \right),  \qquad \gamma^3= \left( \begin{array}{llll} 0 & 0 & 1 & 0 \\
0 & 0 & 0 & -1 \\
-1 & 0 & 0 & 0 \\
0 & 1 & 0 & 0 \end{array} \right) = \left(\begin{array}{ll} 0 &  \sigma^3 \\
- \sigma^3 & 0  \end{array} \right),\nonumber \\
\end{eqnarray}
such that $\gamma^4 =  \gamma^0 \gamma^1 \gamma^2 \gamma^3$, and
the $\sigma^i$
\begin{eqnarray}
&& \sigma^1 = \left(\begin{array}{ll} 0 & 1 \\
1  & 0  \end{array} \right), \qquad \sigma^2 = \left(\begin{array}{ll} 0 & -i \\
i  & 0  \end{array} \right), \qquad \sigma^3 = \left(\begin{array}{ll} 1 & 0 \\
0  & -1  \end{array} \right),
\end{eqnarray}
are the Pauli matrices.

\subsection{Variables separation for the Dirac equation in 5D}

Finally, using the fact that $ \gamma^A = e^A_a \gamma^a$, we
obtain the Dirac equation on the metric (\ref{a1})
\begin{equation}\label{di}
i\, \gamma^A \nabla_A \Psi =0,
\end{equation}
where we shall use (\ref{na}) and (\ref{ga}) and Cartesian
coordinates to describe the 3D Euclidean hypersurface. The Dirac
equation (\ref{di}) can be written as
\begin{equation}
i\, \gamma^a\, e^A_{\,\,a} \,\partial_A \Psi + \frac{i}{8}
\gamma^a\left[\gamma^b,\gamma^c\right] e^A_{\,\,a} e^B_{\,\,b}
\,\,g_{DB} \,\,\left(\partial_A e^D_{\,\,c} + \Gamma_{EA}^D
e^E_{\,\,c}\right) \Psi =0.
\end{equation}
The relevant second kind Christoffel symbols are
\begin{eqnarray}
\Gamma^0_{04} &=& \frac{1}{\psi}, \quad
\Gamma^0_{11}=\Gamma^0_{22}= \Gamma^0_{33} =
\frac{e^{2\frac{t}{\psi_0}}}{\psi_0}, \quad
\Gamma^1_{01} = \Gamma^2_{02}= \Gamma^3_{03}=\frac{1}{\psi_0}, \nonumber \\
\quad \Gamma^1_{14} & = & \Gamma^2_{24}=
\Gamma^3_{34}=\frac{1}{\psi},\quad \Gamma^4_{00} =
\frac{\psi}{\psi^2_0}, \Gamma^4_{11} = \Gamma^4_{22}=
\Gamma^4_{33} = - \frac{e^{2\frac{t}{\psi_0}} \psi}{\psi^2_0}.
\end{eqnarray}
Hence, the Dirac equation on the 5D Riemann-flat metric (\ref{a1})
results to be
\begin{equation}\label{di}
i\,\left\{\gamma^0 \left[\left(\frac{\psi_0}{\psi}\right)
\frac{\partial}{\partial t} + \frac{3}{2 \psi} \right] +
\left(\frac{\psi_0}{\psi}\right) e^{-\frac{t}{\psi_0}} \left[
\vec{\gamma}. \vec{\nabla}\right] + \gamma^4 \left[ \frac{\partial
}{\partial \psi} + \frac{2}{\psi} \right]\right\} \Psi= 0.
\end{equation}
In order to make separation of variables we shall use the method
introduced in \cite{jmp}. After some algebraic manipulation, Eq.
(\ref{di}) can be rewritten as
\begin{equation}
\left(\hat{K}_{04} + \hat{K}_{123}\right) \Phi =0,
\end{equation}
where $\hat{K}_{04} \Phi = k \Phi = - \hat{K}_{123} \Phi$ and
$\Phi = \gamma^0 \gamma^4 \Psi$ [By making $\det|k \Phi
+\hat{K}_{123} \Phi|=0$, we can evaluate the variable separation
constant: $k=|\vec{k}|$]. The operators of separation
$\hat{K}_{04}$ and $\hat{K}_{123}$ are given by
\begin{eqnarray}
&& \hat{K}_{04}= \left[  \gamma^0 e^{t/\psi_0}
\left(\frac{\partial}{\partial t} + \frac{3}{2\psi_0} \right) +
\frac{1}{\psi_0} \gamma^4 e^{t/\psi_0} \left(\psi
\frac{\partial}{\partial\psi} + 2\right) \right] \gamma^0
\gamma^4, \\
&& \hat{K}_{123}= \left[ \gamma^1 \frac{\partial}{\partial x} +
\gamma^2 \frac{\partial}{\partial y} + \gamma^3
\frac{\partial}{\partial z} \right] \gamma^0 \gamma^4,
\end{eqnarray}
where the condition $\left[ \hat{K}_{04}, \hat{K}_{123} \right]
=0$ must be fulfilled. Since the metric (\ref{a1}) is 3D spatially
isotropic, one obtains that the 3D spatial solutions can be
expanded in term of harmonic functions, so that one can write
\begin{equation}
\Phi(t,x,y,z,\psi)\sim \Phi_0(t,\psi)\, e^{i\, \vec{k}.\vec{x}},
\end{equation}
such that $\vec{k}$ is the wavenumber of propagation on the 3D
isotropic and homogeneous Euclidean space. Furthermore, one
obtains
\begin{equation}
\hat{K}_{04} \Phi_0(t,\psi) = k \, \Phi_0(t,\psi).
\end{equation}
After some algebra, we obtain that
\begin{equation}
\left(\hat{K}_0 + \hat{K}_4 \right) \Upsilon(t,\psi) =0,
\end{equation}
such that
\begin{equation}\label{k4}
\hat{K}_4\, \Upsilon = M \Upsilon = - \hat{K}_0 \, \Upsilon,
\end{equation}
where we have used that $\Upsilon(t,\psi) =
\left(\gamma^0\right)^{-1} \, \Phi_0(t,\psi)$, and
\begin{eqnarray}
\hat{K}_0 &= &\left[ \gamma^4 \left(\frac{\partial}{\partial t} +
\frac{3}{2 \psi_0} \right) -
\mathbb{I} \,k\, e^{-t/\psi_0} \right] \gamma^0, \\
\hat{K}_4 & = & \left[ \gamma^0 \frac{1}{\psi_0} \left( \psi
\frac{\partial}{\partial\psi} + 2 \right)\right] \gamma^0,
\end{eqnarray}
that comply with $\left[\hat{K}_0,\hat{K}_4\right]=0$. Using the
first equation in (\ref{k4}), with the variables separation
\begin{equation}
\Upsilon(t,\psi) =\Upsilon_0(t) \, \Lambda(\psi),
\end{equation}
 we obtain the differential equation for
 $\Lambda(\psi)$
 \begin{equation}
 \frac{\psi}{\psi_0} \frac{\partial \Lambda}{\partial \psi} +
 \frac{2}{\psi_0} \Lambda = M \Lambda,
 \end{equation}
 which is the same differential equation obtained in
 \cite{sab}, but in a different framework. The solution for this equation is
\begin{equation}
\Lambda(\psi) = \Lambda_0\left(\frac{\psi}{\psi_0}\right)^{M_0-2},
\end{equation}
where $\Lambda_0$ is a constant of integration and $M=M_0/\psi_0$
is a separation constant. For $M_0 < 2$ the function
$\Lambda(\psi)$ tends to $0$ for $\psi \rightarrow \pm \infty$,
but is divergent for $\psi \rightarrow 0$. In order for the
function $\Lambda(\psi)$ to be real, we must ask $M_0$ to take
integer values and $M_0\leq2$ : $M_0 = 2,1,0,-1,-2,-3,-4, ...$. An
interesting property is that for even $|M_0|$ values the function
$\Lambda(\psi)$ is even but for odd $|M_0|$ values the function is
also odd.

\section{Effective 4D Dirac equation for massive neutrinos in a de
Sitter spacetime}

We assume a static foliation of the 5D spacetime on the 4D
hypersurface $\Sigma_0$, on which the 4D energy momentum tensor is
described by a perfect fluid $\bar{T}_{\alpha\beta}= \left.
e^{A}_{\alpha} e^B_{\beta} T_{AB}\right|_{\psi_0} =(\rho
+P)u_{\alpha}u_{\beta}-P h_{\alpha\beta}$, where
$\rho(t,r,\psi_0)$ and $P(t,r,\psi_0)$ are the energy density and
pressure of the induced matter, respectively. The 4-velocities
$u_{\alpha}$ are related to the 5-velocities $U_A$ by $u_a =
e^A_{\alpha} U_A$, and $h_{\alpha\beta} = e^A_{\alpha} e^B_{\beta}
g_{AB}$ are the components of the metric tensor in (\ref{24}). The
Campbell-Magaard theorem, which is valid in any number of
dimensions, implies that every solution of the 4D Einstein
equations with arbitrary energy momentum tensor can be embedded,
at least locally, in a solution of the 5D Einstein field equations
in vacuum. Because of this, the tensor $\bar{T}_{\mu\nu}$ is
induced as a 4D manifestation of the embedding geometry.

If we take a constant foliation $\psi=\psi_0\neq 0$ [to avoid a
possible divergence of $\Lambda(\psi=\psi_0=1/H_0)$] on the metric
(\ref{a1}) we obtain the metric(\ref{24}), and the second equation
in (\ref{k4}) takes the form
\begin{equation}
\left[ \gamma^4 \gamma^0 \left(\frac{\partial}{\partial t} +
\frac{3 H_0}{2} \right) - \gamma^0 k\, e^{-H_0 t} \right]\,
\Upsilon_0(t) = M \, \Upsilon_0(t),
\end{equation}
where $M=M_0/\psi_0=M_0\,H_0$ is the induced mass of the neutrinos
on the de Sitter spacetime (\ref{24}). If we consider
\begin{equation} \Upsilon_0(t) =
\left(\begin{array}{ll} \Upsilon^{\uparrow}_M(t) \\
\Upsilon^{\downarrow}_M(t)  \end{array} \right),
\end{equation}
such that $\Upsilon^{\uparrow}_M(t)$ and
$\Upsilon^{\downarrow}_M(t)$ comply with the coupled differential
equations
\begin{eqnarray}
\frac{i}{c} \left( \frac{\partial}{\partial t} + \frac{3 H_0}{2}
\right) \Upsilon^{\downarrow}_M + \left( M - k\,
e^{-H_0 t} \right) \Upsilon^{\uparrow}_M & = & 0, \label{c1} \\
-\frac{i}{c} \left( \frac{\partial}{\partial t} + \frac{3 H_0}{2}
\right) \Upsilon^{\uparrow}_M + \left( M + k\, e^{-H_0 t} \right)
\Upsilon^{\downarrow}_M & = & 0. \label{c2}
\end{eqnarray}
One can work with both coupled Eqs. (\ref{c1}) and (\ref{c2}) in
order to decouple $\Upsilon^{\uparrow}_M(t)$ and
$\Upsilon^{\downarrow}_M(t)$, and obtain two decoupled second
order differential equations
\begin{eqnarray}
\frac{\partial^2 \Upsilon^{\uparrow}_M}{\partial t^2} &+&
\frac{H_0 \left(3M + 4 k\, e^{-H_0 t }\right)}{\left(M + k\, e^{-
H_0t }\right)} \frac{\partial
\Upsilon^{\uparrow}_M}{\partial t} \nonumber \\
&+& \left[\frac{3\, H^2_0 \,k\, e^{-H_0 t}}{ 2  \left(M + k
\,e^{-H_0 t}\right)} + \frac{9 H^2_0}{4} - \left( M^2 - k^2
e^{-2H_0
t}\right) \right]\, \Upsilon^{\uparrow}_M  =  0, \label{ec1} \\
\frac{\partial^2 \Upsilon^{\downarrow}_M}{\partial t^2} &+&
\frac{H_0 \left(3M - 4 k\, e^{-H_0 t}\right)}{ \left(M - k\,
e^{-H_0 t}\right)} \frac{\partial
\Upsilon^{\downarrow}_M}{\partial t} \nonumber \\
&+& \left[-\frac{3\, H^2_0 \,k\, e^{-H_0 t}}{ 2 \left(M - k
\,e^{-H_0 t}\right)} + \frac{9 H^2_0}{4} - \left( M^2 - k^2
e^{-2H_0 t}\right) \right]\, \Upsilon^{\downarrow}_M = 0.
\label{ec2}
\end{eqnarray}
The general solutions of Eqs. (\ref{ec1}) and (\ref{ec2}),
are\footnote{In the case of $M=0$, both solutions
$\Upsilon^{(\uparrow, \downarrow)}_{M=0}(t)$, are equal and the
general solution takes the particular form
\begin{equation}
\Upsilon^{(\uparrow, \downarrow)}_{M=0}(t) = e^{-\frac{3}{2} H_0
t} \left\{ {\rm C_1} \sin \left[ \frac{k}{H_0}\, e^{-H_0 t}
\right] + {\rm C_2} \, \cos \left[\frac{k}{H_0}\, e^{-H_0 t}
\right] \right\}.\nonumber
\end{equation}}
\footnote{From the structure of the eq. (\ref{di}) one can see
that $\Upsilon^{(\uparrow, \downarrow)}_{M}$ are not Pauli
spinors.}
\begin{eqnarray}
\Upsilon^{(\uparrow, \downarrow)}_M(t) & = & e^{\mp\frac{i}{H_0} k
e^{-H_0 t}} \left\{ {\rm C_1} \, e^{-\left[ \frac{3}{2}+ {M\over
H_0} \right] {H_0} t} \, {\cal H}{\rm eunC}
\left[a \,i,a,-2,0,1, \mp x(t) \right] \right. \nonumber \\
& +& \left. {\rm C_2} \, e^{-\left[ \frac{3}{2}- {M\over H_0}
\right] {H_0} t} \, {\cal H}{\rm eunC} \left[a \,i,-a,-2,0,1, \mp
x(t) \right]\right\}, \label{sol}
\end{eqnarray}
where ${\cal H}{\rm eunC} \left[a \,i,a,-2,0,1, \mp x(t) \right]$
and ${\cal H}{\rm eunC} \left[a \,i,-a,-2,0,1, \mp x(t) \right]$
are the Heun Confluent functions with arguments $\mp x(t)= \mp
\frac{k}{M} \, e^{-H_0 t}$, and parameters ($a\,i,\mp a,-2,0,1$).
Here
\begin{eqnarray}
a & = & 2 \,\frac{M}{H_0} .\nonumber
\end{eqnarray}
Notice that for late times the Heun functions become
\begin{eqnarray}
\left. {\cal H}{\rm eunC} \left[a\,i,a,-2,0,1, \mp x(t)
\right]\right|_{H_0 t \gg 1} \rightarrow 1,  \\
\left. {\cal H}{\rm eunC} \left[a\,i,-a,-2,0,1, \mp x(t)
\right]\right|_{H_0 t \gg 1} \rightarrow \infty.
\end{eqnarray}
Furthermore, in this situation one obtains that
\begin{eqnarray}
&& \left. e^{\mp\frac{i}{H_0} k^{3/2} e^{-H_0 t}}\right|_{H_0 t \gg 1} \rightarrow 1, \\
&& \left.e^{-\left[ 9+\sqrt{16 {M^2\over H^2_0} + 9 } \right]
\frac{H_0}{4} t}\right|_{H_0 t \gg 1} \rightarrow 0.
\end{eqnarray}
In order for the spinors $\Upsilon^{(\uparrow, \downarrow)}_M$ to
be well behaved for late times of inflation, the argument $\left[
\frac{3}{2}+{M\over H_0} \right]$ of the exponential $e^{-\left[
\frac{3}{2}+{M\over H_0} \right] {H_0} t}$ must be positive or
zero, and $\rm C_2=0$. This requirement only fix a lower bound on
$M_0$, so that $-\frac{3}{2}\leq M_0\leq2$. There are three
possible values of mass, which are
\begin{equation}
M_1=0, \qquad M_2 =  H_0, \qquad M_3= 2 H_0.
\end{equation}
This is a very strong result. Notice that if we make the
extrapolation to present day values of the Hubble parameter,
$\bar{H_0} \simeq 10^{-33} \,H_0$, one would obtain that the
present day masses of the neutrinos are
\begin{equation}
\bar{M}_1 =0, \qquad \bar{M}_2 = \bar{H_0} \times 10^{-12}\, eV,
\qquad \bar{M}_3 = \bar{H}_0 \simeq 2 \times 10^{-12}\, eV,
\end{equation}
where we have taken $H_0 \simeq 10^{-9} G^{-1/2}$. These results
are in agreement with evidence\cite{rmp}.

Finally, for very large times the solution (\ref{sol}) can be
approximated to
\begin{equation}
\left. \Upsilon^{(\uparrow, \downarrow)}_M(t)\right|_{ H_0 t \gg
1} \simeq {\rm C_1}\, e^{\mp\frac{i}{H_0} k e^{-H_0 t}}\,
e^{-\left[ \frac{3}{2}+{M\over H_0} \right] {H_0} t}.
\end{equation}
where the spinors are normalized by $\left<\Psi | \Psi\right> =
1$.\\

\section{Final Remarks}

We have studied the Dirac equation and solutions for effective 4D
massive neutrinos on a de Sitter expansion, from a 5D Riemann-flat
(and hence Ricci-flat), extended de Sitter metric. On this metric
we have defined a 5D vacuum to test massless non-interacting
fermion fields which are minimally coupled to gravity. After
making a static foliation these fermions acquire an induced mass
on the effective 4D curved de Sitter spacetime. However, the mass
of the neutrinos can take only three possible values ($M_1=0 $,
$M_2= H_0 =1/\psi_0$ and $M_3= 2/\psi_0$). If we consider present
day values of the Hubble parameter the bigger mass $\bar{M}_2$
should be close to $\sim 2 \times 10^{-12}\, {\rm eV}$. This is a
very strong result which assures that the effective 4D masses of
the neutrinos are inversely proportional to the foliation,
$\psi=\psi_0$, and shows how the mass of the neutrinos on a 4D de
Sitter spacetime can be induced from a free massless 5D test
spinors on a extended Riemann-flat (and hence Ricci-flat) metric
which has non-zero connections $\Gamma^a_{bc}$. But more strong is
the result that the masses of cosmological neutrinos should be
dependent on the energy scale of the universe because they are
dependent on the Hubble parameter. However, these results must be
taken carefully because we have neglected any kind of interactions
of the spinor fields with other fields. (These results are only
valid for free 4D neutrinos that propagate freely in a 4D de
Sitter background.) A more complete treatment will be studied in a
future work.

\section*{Acknowledgements}

\noindent The authors acknowledge UNMdP and CONICET Argentina for
financial support.

\end{document}